\documentclass[conference]{IEEEtran}
\IEEEoverridecommandlockouts
\usepackage{cite}
\usepackage{amsmath,amssymb,amsfonts}
\usepackage{algorithmic}
\usepackage{graphicx}
\usepackage{textcomp}
\usepackage{xcolor}
\def\BibTeX{{\rm B\kern-.05em{\sc i\kern-.025em b}\kern-.08em
    T\kern-.1667em\lower.7ex\hbox{E}\kern-.125emX}}
    
\usepackage{subfigure}
\usepackage{tikz}
\usepackage{graphicx}
\usepackage{braket}

\begin{document}

\title{Towards solving large QUBO problems using quantum algorithms: improving the LogQ scheme 
}

\author{\IEEEauthorblockN{1\textsuperscript{st} Yagnik Chatterjee}
\IEEEauthorblockA{
\textit{TotalEnergies}\\
2, place Jean Millier\\
92078 Paris La Défense Cedex, France\\
yagnik.chatterjee@totalenergies.com}
\and
\IEEEauthorblockN{2\textsuperscript{nd} Jérémie Messud}
\IEEEauthorblockA{
\textit{TotalEnergies}\\
2, place Jean Millier\\
92078 Paris La Défense Cedex, France\\
jeremie.messud@totalenergies.com}
}
\maketitle

\begin{abstract}
The LogQ algorithm encodes Quadratic Unconstrained Binary Optimization (QUBO) problems with exponentially fewer qubits than the Quantum Approximate Optimization Algorithm (QAOA).
The advantages of conventional LogQ are accompanied by a challenge related to the optimization of its free parameters, which requires the usage of resource-intensive 
evolutionary or even global optimization algorithms.
We propose a new LogQ parameterization that can be optimized with a gradient-inspired method, which is less resource-intensive and thus strengthens the advantage of LogQ  over QAOA for large/industrial problems.
We illustrate the features of our method on an analytical model and present larger scale numerical results on MaxCut problems.
\end{abstract}


\section{Introduction}

Finding exact solutions to Quadratic Unconstrained Binary Optimization (QUBO) problems represents an active space of research \cite{qubotutorial, date2021qubo, calude2017qubo, papalitsas2019qubo}, which would benefit many industrial applications (combinatorial optimization of portfolios, fleet, charging stations, etc.).
These problems are NP-hard \cite{karp, surveypoly} for classical computers. The promise of Quantum Computing (QC) lies in its potential to reach the solutions or at least provide better approximations compared to classical heuristics in a reasonable time.
Hybrid quantum-classical algorithms have been intensively studied for such tasks \cite{cerezo2021variational,peruzzo2014variational,moll2018quantum}, the Quantum Approximate Optimization Algorithm (QAOA) probably being the most popular \cite{qaoa,qaoa2}. The number of qubits required for QAOA scales linearly with problem size and the depth of the QAOA quantum circuit  (number of CNOTs) scales polynomially. This means that a QUBO problem with $n$ binary variables to optimize would require a quantum processing unit (QPU) with $n$ sufficiently low error-rate qubits that can handle quantum circuits of depth polynomial in $n$, which is challenging in the Noisy Intermediate-Scale Quantum (NISQ) era especially for industrial applications (large $n$).

It is therefore worth considering algorithms that would reduce the number of required qubits and the depth of quantum circuits.
The LogQ scheme was proposed as a step in this direction \cite{rancic, chatterjee2023solving}. It is based on a kind of amplitude encoding of QUBO-style problems and requires $\lceil\log n\rceil$ qubits to deal with $n$ QUBO variables ($\lceil.\rceil$ stands for the ceiling function),
i.e. exponentially less qubits than QAOA. In addition, the depth of the quantum circuit (number of CNOTs) 
scales linearly in $n$.
These properties, summarized in Table \ref{complexity_analysis}, seem to make LogQ more compliant and scalable than QAOA for industrial applications \cite{chatterjee2023solving, chatterjee2023hybrid, neasqc}.

However, the advantages of LogQ are accompanied by two challenges: one related to the number of measurements or decomposition of the LogQ operator (which scales at most quadratically in $n$), and the other to the free parameters optimization. Research is ongoing on the first challenge 
but it does not hamper LogQ from remaining competitive in a wide range of situations.
We will come back to this point in the conclusion.
The second challenge is more problematic: the current formulation prevents the usage of a gradient-inspired optimization of the parameters due to a vanishing gradient issue, requiring the usage of evolutionary or even global optimization methods that are resource-intensive. 

We propose a new LogQ parameterization that leads to a gradient-compliant objective function landscape and thus can be optimized with a gradient-inspired scheme, making the parameters optimization much less resource intensive.
We justify the method on an analytical model. We then numerically evaluate its advantages on MaxCut problems, a type of QUBO problem. This work represents a step towards our goal of applying LogQ to industrial problems such as portfolio optimization.
\begin{table}[htbp]
\caption{Comparison between LogQ and QAOA for a QUBO problem with $n$ variables. $p$ represents the number of QAOA layers.}
\centering
    \begin{tabular}{|c|c|c|}
        \hline
         & LogQ & QAOA  \\
        \hline
        No. qubits & $\lceil \log_2 n \rceil=N$ & $n$ \\
        \hline
        No. param. to optimize & $n$ & $2p$  \\
        \hline
        No. CNOTs & $n$ & $p(n^2-n)$  \\
        \hline
    \end{tabular}
    \label{complexity_analysis}
\end{table}

\section{MaxCut and QAOA}

Let us consider an undirected and weighted graph
$(V,E,w)$ with $|V|=n$ the number of vertices, $|E|\leq (n^2-n)/2$ the number of edges and $w_{ij}$ the weights on the edges.
The MaxCut problem aims to resolve
\begin{align}
\label{eq:maxcut}
x^*=\underset{x \in \{1,-1\}^{n}}{\text{argmax}}\frac{1}{2}\sum_{(i,j)\in E} w_{ij} (1-x_ix_j)
,
\end{align}
which provides a partition of the graph vertices into 2 subsets (those with $x_i^*=1$ and those with $x_i^*=-1$) such that the sum of the weights of edges connecting the two subsets is maximized. 
This problem has applications in many industrial fields 
but the solution space grows exponentially with $n$ (in $2^n$), 
making it infeasible for classical computers when $n$ is large.



QAOA is one of the most widely studied quantum algorithms for the MaxCut problem.
The idea is to embed the problem into a Hamiltonian  quantum operator acting on a parameterized $n$-qubit state $\ket{\varphi(\Theta)}$:
\begin{align}
\hat{H} &= - \frac{1}{2}\sum_{(i,j) \in E} w_{ij}(1-Z_i \otimes Z_j),
\end{align}
where $Z_i$ denotes the $Z$ Pauli operator acting on qubit $i$. 

The ground state of $\hat{H}$ encodes the MaxCut solution
through computational basis state encoding. 
This means that the $2^n$ bitstrings related to an $n$-vertex graph problem solution space are encoded in the $2^n$ computational basis states of the $n$-qubit Hilbert space,
and that the ground state of $\hat{H}$ equals the particular basis states related to the MaxCut solution bitstring.
A $Z$ basis measurement of the ground state 
should thus give the MaxCut solution bitstring with probability $O(1)$. 

In practice, things are more complicated. A parameterization is chosen for $\ket{\varphi(\Theta)}$ with the hope that there exist an optimum value $\Theta^*$ for which $\ket{\varphi(\Theta^*)}$ approximates the ground state of $\hat{H}$.
The QAOA parameterization is inspired from the adiabatic theorem.
QAOA starts from the ground state of a trial Hamiltonian and evolves it towards the ground state of $\hat{H}$ in $p$ steps (or $p$ layers), requiring the optimization of $2p$ parameters $\Theta=(\Theta_0,\dots,\Theta_{2p-1})$.
In practice, $p$ is chosen heuristically (often linearly in $n$\cite{brandhofer2022benchmarking,fuchs2021efficient}) and $\Theta^*$ is found by classical optimization,  hoping that the highest probability measurement of $\ket{\varphi(\Theta^*)}$ gives the solution to the MaxCut problem.
The number of CNOT gates required is equal to $p|E|\le p(n^2-n)/2$. 


\section{Laplacian MaxCut version and LogQ}
\label{sec:Laplacian}

A graph Laplacian matrix is defined as
\begin{equation}
\label{laplacian}
L_{ij}=\begin{cases}
          degree(i) \quad &\text{if } \, i=j  \\
          -w_{ij} \quad &\text{if } \, i \neq j \text{ and } (i,j) \in E \\
          0 \quad &\text{otherwise},
     \end{cases}
\end{equation}
where $degree(i)$ refers to the 
sum of the weights of all edges containing the vertex $i$.
The MaxCut problem (\ref{eq:maxcut}) can equivalently be written $x^*=\text{argmax}_{x \in \{1,-1\}^{n}}\frac{1}{4}x^T L x$.

A lighter formulation than QAOA is obtained using the quantum equivalent of the
Laplacian matrix (instead of the Hamiltonian matrix) and amplitude encoding (instead of computational basis state encoding) \cite{rath2023quantumdataencodingcomparative}, through:
\begin{equation}
\label{eqn2}
    \theta^*=\underset{\theta}{\text{argmin }} C(\theta)
    \begin{cases}
          C(\theta)=-2^{N-2}\bra{\Psi(\theta)}\hat{L}\ket{\Psi(\theta)}  \\
          \ket{\Psi(\theta)} = \frac{1}{2^N}\sum_{z=0}^{2^N-1}e^{i \pi R(\theta_z)}\ket{z},
     \end{cases}
\end{equation}
where $N=\lceil\log_2n \rceil$ ($\lceil.\rceil$ stands for the ceiling function), $\ket{\Psi(\theta)}$ is a $N$-qubit state, 
$\{\ket{z}, z=0\dots 2^N-1\}$ are the computational basis states,
$\theta=(\theta_0\dots\theta_{2^N-1})$
are parameters to optimize classically, $e^{i \pi R(\theta_z)}$ are 'amplitudes' that aim to encode the solution to the MaxCut problem
and $\hat{L}$ is the Laplacian quantum operator (that is Hermitian thus an observable). A straightforward method to compute the latter is to perform the decomposition
\begin{equation}
\label{L_decompo}
\hat{L}=\dfrac{1}{2^N}\sum\limits_{k=1}^{4^N}Tr(J_k L)J_k,
\end{equation}
where the $J_k$ represent the $4^N$ possible tensor products of $N$ Pauli matrices and the identity matrix~
\footnote{When $n$ is not an exact power of 2, null matrices of size $2^N-n$ are used to adjust $\hat{L}$ to size $2^N\times 2^N$.}.
The evaluation of the expectation value of $\hat{L}$ then requires at most $(4^N+2^N)/2\approx (n^2+n)/2$ measurements on the Pauli basis (as $\hat{L}$ is a symmetric matrix).

All that remains is to describe the amplitude encoding function $R$ in (\ref{eqn2}), which is constrained as:
\begin{align}
\label{con_1}
& R(\theta_z)\in[0,1]\quad \&\quad
R(\theta_z)\text{ }
reaches\text{ }0\text{ } and\text{ }1\text{ }on\text{ }[0,2\pi]
\\
& R(\theta_z^*)=0\text{ }or\text{ }1\text{ }(only)\text{ }once\text{ }\theta\text{ }parameters\text{ }optimized
.\nonumber
\end{align}
Then, the MaxCut solution is encoded in $R(\theta_z^*)$: for each vertex $z$ of the graph, $R(\theta_z^*)=0$ (resp. $1$) means that the vertex belongs to the 1$^{st}$  (resp. the 2$^{nd}$) MaxCut set \cite{chatterjee2023solving}.
Interestingly, 
each amplitude of the $2^N\approx n$ computational basis states in $\ket{\Psi(\theta^*)}$, equation (\ref{eqn2}), then encodes only two opposite values ($\pm 1$ related to $R(\theta_z^*)=0$ or $1$), making the scheme a 'light' amplitude encoding that should be robust to noise.

The main advantage of the scheme is to allow the treatment of MaxCut and, more generally, QUBO problems with exponentially fewer qubits than QAOA, i.e. with $N=\lceil\log_2n \rceil$ qubits, hence the name LogQ. 
The scheme requires $2^N\approx n$ CNOTs, i.e. polynomially fewer than QAOA, see \cite{chatterjee2023solving} for a discussion.
These properties are appealing for usage on large/industrial optimization problems.
Let us recall Table \ref{complexity_analysis}, which compares LogQ and QAOA complexities.
\begin{figure}[!h]
\begin{center}
  \includegraphics[scale=0.412]{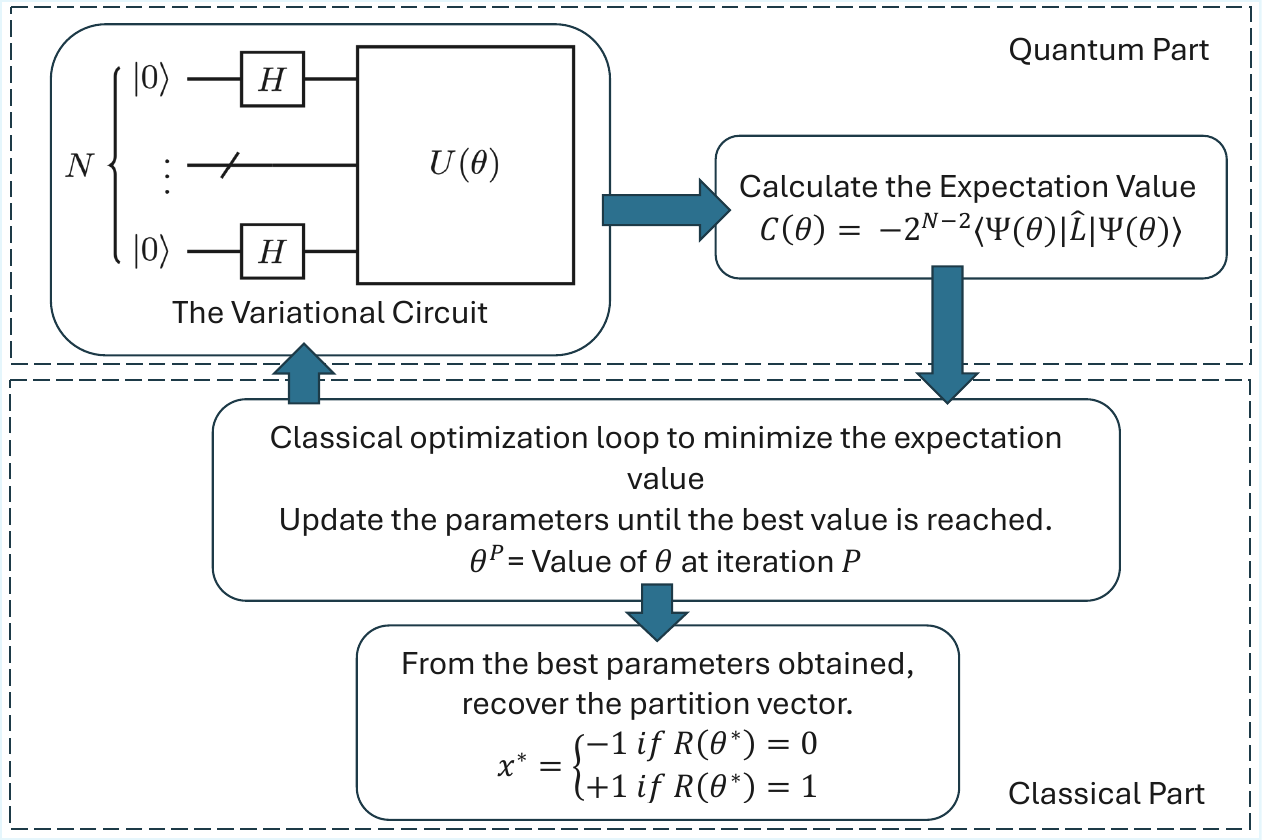}
  \caption{Diagrammatic representation of the hybrid quantum-classical LogQ algorithm using the notations of section \ref{sec:Laplacian}.}
  \label{algprocess}
  \end{center}
\end{figure}

Finally, note that the $N$-qubit state in (\ref{eqn2}) can equivalently be written as
\begin{align}
\label{U}
    \ket{\Psi(\theta)}&=U(\theta)H^{\otimes N} \ket{0}^{\otimes N}
    \\
    U(\theta)&=
    \text{diag}\Big[e^{i \pi R(\theta_0)},\dots,e^{i \pi R(\theta_{2^N-1})}\Big],
    \nonumber
\end{align}
where $U(\theta)$ is a unitary operator, which is useful to implement LogQ as a quantum circuit on a QPU. 

LogQ is a hybrid quantum-classical algorithm as the optimization 
of the $\theta$ parameters in (\ref{eqn2})  is performed using classical methods.
The LogQ algorithm is schematized in Fig. \ref{algprocess}.

Defining $R$ represents the only heuristics of the LogQ scheme and is the focus of the next sections.
We first develop various constraints on $R$, and then
explore different parameterizations compatible with these constraints and their implications regarding the type of classical optimization performed within LogQ.
Our goal is to release the need of evolutionary or even global optimization methods which are required within the original LogQ scheme,
and develop a gradient or local optimization compliant version which would be much-less resource intensive.

\section{Additional constraints on $R$}

%
The second line of (\ref{con_1}) constrains $R$ to equal 0 or 1 (only) once the parameters are optimized. 
Considering $R$ (and thus $C$) is differentiable
almost everywhere and that optimal parameters lead to a null cost function gradient,
the discussed constraint can be enforced through the condition:
\begin{align}
\label{con_interm}
    \partial C(\theta)/\partial \theta_z =0
    \quad\Rightarrow\quad
    R(\theta_z)=0\text{ }or\text{ }1
    .
\end{align}
%
%
Using (\ref{eqn2}) and the fact that $\hat{L}$ is Hermitian, we have 
\begin{align}
\label{eqn_c_valley}
    C(\theta)\propto& \sum_{z=0}^{2^N-1}
    \bra{z}\hat{L}\ket{z}
\\    
    -2&\sum_{z>w=0}^{2^N-1}
    \cos \big[\pi (R(\theta_w)-R(\theta_z))\big]
    \bra{w}\hat{L}\ket{z}
    ,
    \nonumber
\end{align}
and thus
\begin{align}
     \frac{\partial C(\theta)}{\partial \theta_z}\propto 2 R'(\theta_z)\sum_{z>w=0}^{2^N-1}\sin\big[ \pi(R(\theta_{w})-R(\theta_z)) \big]
    \bra{w}L\ket{z}.
\label{eqn_grad}
\end{align}
From the latter equation, we can deduce that a necessary condition for (\ref{con_interm}) to hold is:
%
\begin{align}
& R'(\theta_z)=0 \quad\Rightarrow  \quad 
R(\theta_z)=0 \text{ }or \text{ } 1,
\nonumber\\
& (R(\theta_z)\text{ }differentiable\text{ }almost\text{ }everywhere.)
\label{con_2}
\end{align}
Of course (\ref{con_interm}) is more general than (\ref{con_2}), but the interest of (\ref{con_2}) is to give an explicit constraint on $R$.

%
What further constraints do we have on $R$?
Despite is advantages, the LogQ scheme can be efficient only if we can define a $R$ that allows efficient parameter optimization, ideally using gradient-inspired methods as the latter have low-order polynomial complexities in $n$. 
For gradient-inspired methods to work, $\partial C(\theta)/\partial\theta$ must not vanish in a large part of the domain and any vanishing gradient must be related to the cost function reaching its maximum or minimum value
(these values can be reached many times but local minima are forbidden).
In this direction and considering (\ref{eqn_c_valley})-(\ref{eqn_grad}), we propose two additional conditions on $R$:
\begin{align}
&\hspace{-9mm}There\text{ }exist\text{ }large\text{ }intervals\text{ } in\text{ }[0,2\pi]\text{ }such\text{ }that
\nonumber\\
&\hspace{-8mm}\text{ }\text{ }|R'(\theta_z)|\text{ }is\text{ }not\text{ }small.
\label{con_3}
\end{align}
and
\begin{align}
&\forall\theta_z\text{ }such\text{ }that\text{ }|R'(\theta_z)|\text{ }is\text{ }small:
\nonumber\\
&\text{ }\text{ }\text{ }C(\theta) \text{ }is\text{ }either\text{ }close\text{ }to\text{ } \max_{\theta_z} C(\theta) \text{ }or\text{ }\min_{\theta_z} C(\theta).
\label{con_4}
\end{align}
These are intentionally loosely formulated conditions as, 
in practice, '$large$, '$small$' and '$close$' will depend
on the gradient-inspired method used. Methods using gradient perturbations (stochastic or others) allow us to escape 'small' local minima, i.e. to 'jump' above small 'barriers' in the cost function valley, and to escape 'small' plateaus, i.e. areas where the gradient tends to vanish.
Gradient-free methods like Cobyla also allow to some extent to escape local minima and plateaus (they can be considered as methods that compute a kind of gradient perturbation in case of differentiable functions).
Typically, a large 'rhobeg' parameter in Cobyla (determining the size of the 'trust region' around a point) allows to somewhat control these aspects.
A consequence is that such optimization methods provide some flexibility
on the '$large$, '$small$' and '$close$' in (\ref{con_3})-(\ref{con_4}).
By abuse of language, we call all of them 'gradient-inspired' methods in the following.


\section{Issue with Original LogQ and way forward}

The original LogQ \cite{chatterjee2023solving} scheme mostly uses 
    \begin{equation}
    \label{eqn_grad2}
        R\rightarrow R^{(0)}(\theta_z)=
          \begin{cases}
          0\quad &\text{if } \,  \theta_z \in [0,\pi[ \\
          1 \quad &\text{if } \,  \theta_z \in [\pi,2\pi],
          \end{cases}
    \end{equation}
which satisfies constraints (\ref{con_1}), (\ref{con_2}) and (\ref{con_4})
but not (\ref{con_3}).
Indeed, $R'(\theta_z)$ being null almost everywhere, the scheme is not optimizable using gradient-inspired methods.
Another parameterization has been proposed in \cite{rancic}, but it also does not fully satisfy (\ref{con_3}) and leads to parameter optimization issues due to the highly oscillating cost function valley it produces.
As a consequence, all previous implementations of LogQ based on these $R$ used Genetic Algorithms (GA) to optimize the parameters \cite{rancic, chatterjee2023solving}. This hampers the LogQ efficiency as GAs
are more costly than gradient-inspired algorithms, usually scaling highly polynomially (and even exponentially) with $n$ to achieve very good accuracy. 
The question of finding $R$ parameterizations that would allow us to use gradient-inspired methods is still open. 

\begin{figure}[ht]
\begin{center}
  \includegraphics[scale=0.55]{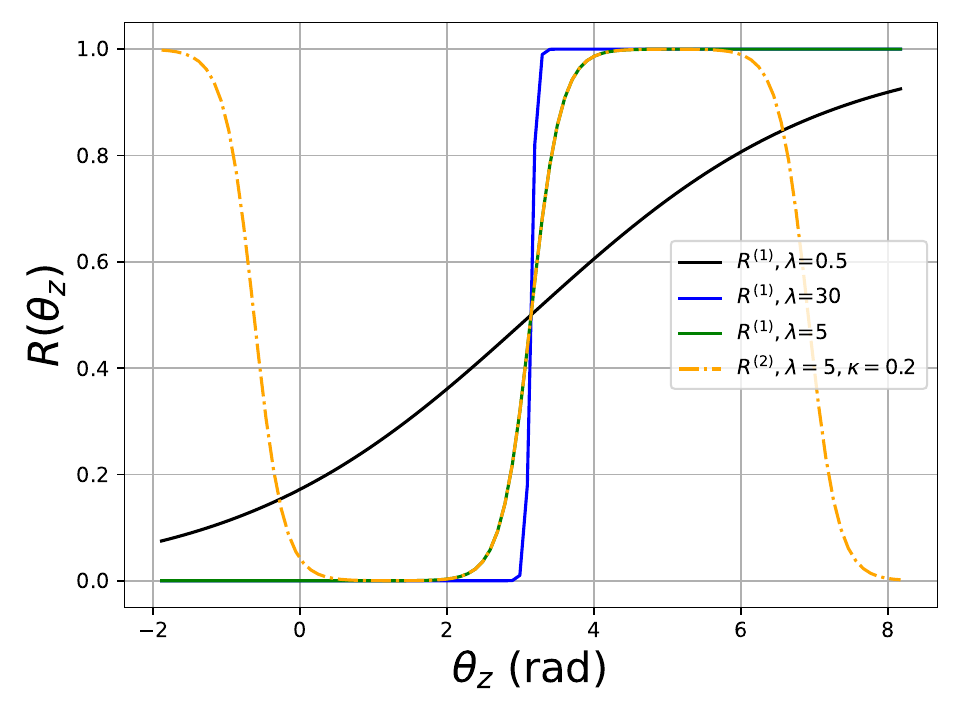}
  \caption{Representation of $R$ given by (\ref{eqn_grad3}) for $\lambda=30$, $5$, $0.5$, and of $R$ given by (\ref{eqn_grad4}) for $\lambda=5$, $\kappa=0.2$, in $[-0.6\pi,2.6\pi]$.}
  \label{fig_R}
  \end{center}
\end{figure}

A natural choice to generalize (\ref{eqn_grad2}) is the sigmoid function \cite{yang2018efficient},
whose smoothness is controlled by a parameter $\lambda>0$ ($\lambda\rightarrow+\infty$ leading to (\ref{eqn_grad2})):
    \begin{equation}
    \label{eqn_grad3}
        R\rightarrow R^{(1)}_\lambda(\theta_z)=\frac{1}{1+e^{\lambda(\pi-\theta_z)}}=\text{sgm}_\lambda(\pi-\theta_z).  
    \end{equation}
Taking a well-chosen $\lambda$ value 
and considering $\theta_z\in[0,2\pi]$
allows us to satisfy the constraints  (\ref{con_1}), (\ref{con_2}) and (\ref{con_3}) but no longer (\ref{con_4}) (local minima appear as will be illustrated later).
A more suitable form consists of the following distortion of the sigmoid on the edges of the interval
$[-\gamma\pi,(2+\gamma)\pi]$:
    \begin{align}
    \label{eqn_grad4}
        R\rightarrow R^{(2)}_{\lambda,\kappa}(\theta_z)=&
          \text{sgm}_\lambda(\pi-\theta_z)\times\text{sgm}_{-\lambda}(\kappa\pi+2\pi-\theta_z)
\nonumber\\
          &+\text{sgm}_{\lambda}(\kappa\pi-\theta_z),
    \end{align}
for $0\le\kappa<\gamma\le 1$.
Fig. \ref{fig_R} gives an illustration of $R^{(1)}_\lambda$ 
and $R^{(2)}_{\lambda,\kappa=0.2}$ 
for various $\lambda$ values, on the $[-0.6\pi,2.6\pi]$ interval ($\gamma=0.6$).

In the next section, we will illustrate why it is efficient in practice to slightly extend the range of $\theta_z$ beyond $[0,2\pi]$ using $\gamma>0$.
We will also show, using an analytical model, that $R^{(2)}_{\lambda,\kappa=0.2}$
enables the satisfaction of all constraints developed above
to an extent that is reasonable for a customized gradient-inspired method to work.


\section{Analytical model and learning on $R$}

\begin{figure}[ht]
\begin{center}
    \begin{tikzpicture}
    \foreach \x/\color in {0/red, 1/white, 2/red, 3/white} {
        \node[circle, draw, fill=\color] (\x) at (360/6*\x:2.9cm) {$\x$};
    }
    \draw (0) -- node[pos=0.5, above] {$3$} (1);
    \draw (1) -- node[pos=0.5, above] {$8$} (2);
    \draw (2) -- node[pos=0.5, left] {$4$} (3);
    \draw (0) -- node[pos=0.5, below] {$1$} (2);
    \end{tikzpicture}
    \caption{Graph and the two sets of the MaxCut problem solution (each set contains the vertices with same color).}
  \label{fig_graph}
  \end{center}
\end{figure}
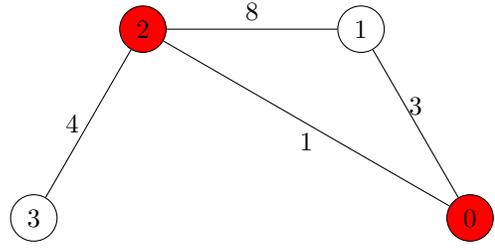

Let us consider a $n=4$ vertices graph $(V,E,w)$
with $V \in \{0,1,2,3\}$, $E\in \{(0,1), (0,2), (1,2), (2,3)\}$ and $w\in \{3,1,8,4\}$,  represented in Fig. \ref{fig_graph}. The MaxCut solution sets contain the vertices with the same color and the cut value is equal to 15 (=4+8+3).

We need $N=\log_2 n=2$ qubits in the LogQ scheme. The $2^N=4$ corresponding computational basis states are $\{\ket{z}, z=0\dots3\}$ 
$=\{\ket{00},\ket{01},\ket{10},\ket{11}\}$.
The LogQ state
$\ket{\Psi(\theta)}\propto[e^{i \pi R(\theta_0)},e^{i \pi R(\theta_1)},e^{i \pi R(\theta_2)},e^{i \pi R(\theta_3)}]^t$
contains $2^N=4$ parameters to optimize. 
Using  (\ref{laplacian}), we compute
the Laplacian matrix of the graph:
\begin{equation}\label{example_laplacian}
    L=\begin{bmatrix}
        4 & -3 & -1 & 0  \\
        -3 & 11 & -8 & 0 \\
        -1 & -8 & 13 & -4   \\
        0 & 0 & -4 & 4
    \end{bmatrix}.
\end{equation}
$L$ must be converted  into a sum of Pauli string operators for a QPU implementation
\footnote{
$\hat{L} = 8I\otimes I - 3.5I\otimes X + 0.5I\otimes Z
- 0.5X\otimes I - 4X\otimes X - 0.5 X\otimes Z
- 4Y\otimes Y -0.5 Z\otimes I + 0.5Z\otimes X - 4Z\otimes Z$.
},
but this is not mandatory for the considerations in the following. Using (\ref{eqn_c_valley}) we obtain:
\begin{align}
\label{eq_C}
    C(\theta)=& 1.5\cos(\pi [R(\theta_1)-R(\theta_0)])\\
        &  +0.5\cos(\pi [R(\theta_2)-R(\theta_0)])\nonumber \\
        &+4\cos(\pi [R(\theta_2)-R(\theta_1)])\nonumber\\
        &+2\cos(\pi [R(\theta_3)-R(\theta_2)])-8.\nonumber
\end{align}

\begin{figure}[h] 
    \centering
        \includegraphics[scale=0.18]{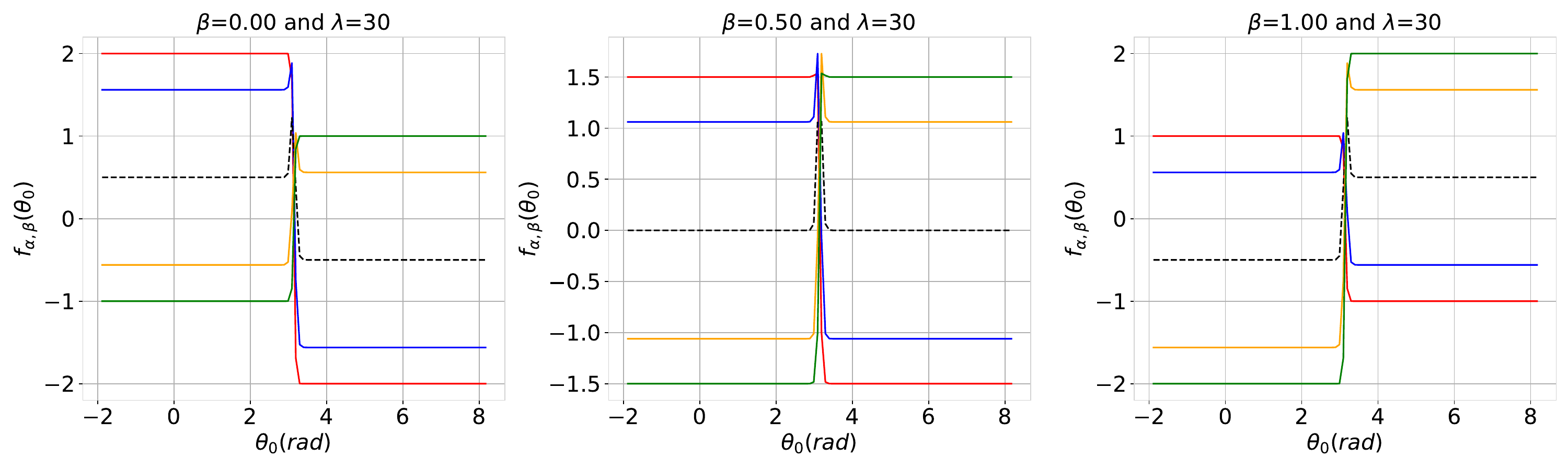}
        \includegraphics[scale=0.18]{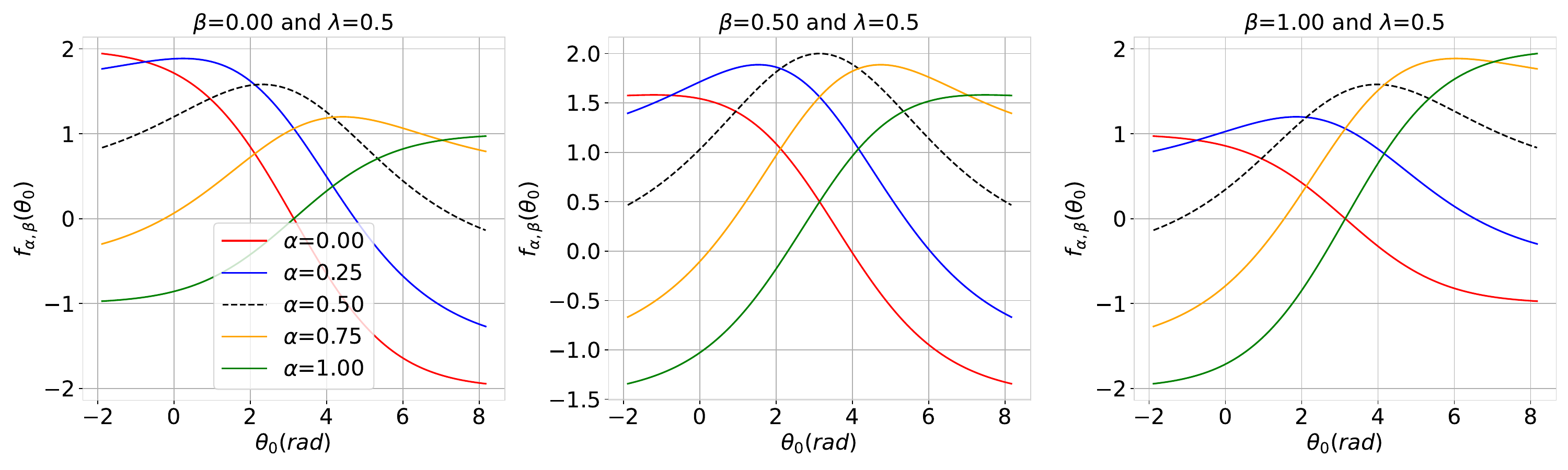}
        \includegraphics[scale=0.18]{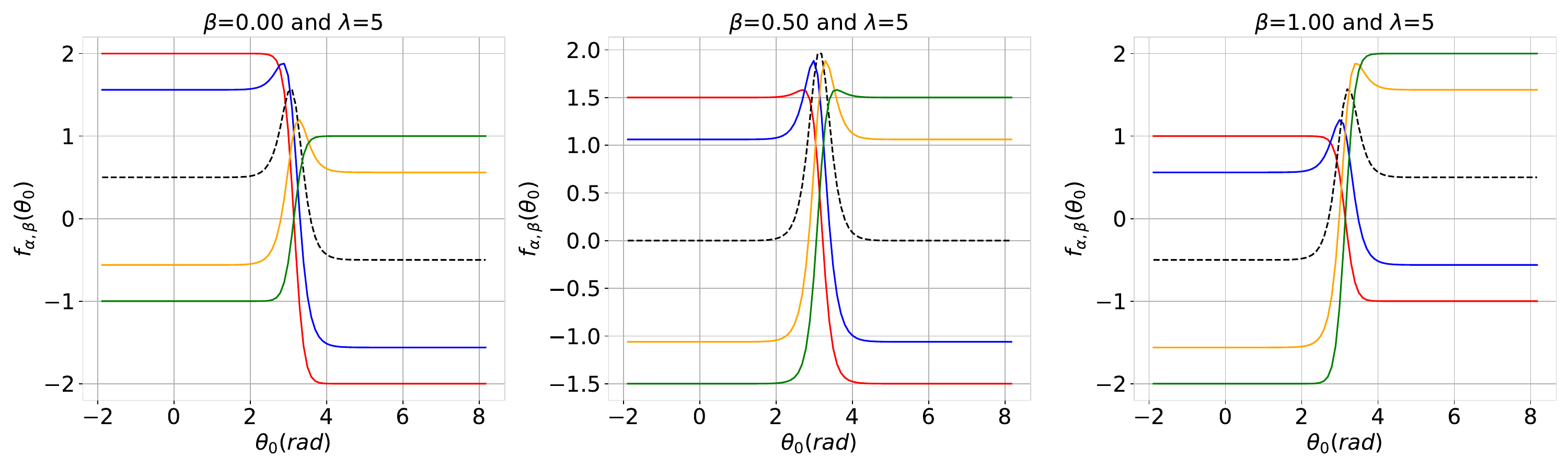}
    \caption{Representation of $f_{\alpha,\beta}(\theta_0)$ using $R$ given by (\ref{eqn_grad3}) for $\lambda=30$ (top row), $\lambda=0.5$ (middle row), $\lambda=5$ (bottom row), and $\theta_0\in[-0.6\pi,2.6\pi]$. Colors: different $\alpha$ values. Columns: different $\beta$ values.}
    \label{cost_ft_valley}
\end{figure}

We can now study the behavior of the cost function valley,
e.g. the $\theta_0$-direction variations of $C(\theta)$ described by
\begin{align}
    f_{\alpha,\beta}(\theta_0) =& 1.5\cos(\pi [\alpha-R(\theta_0)])\\
    &+0.5\cos(\pi [\beta-R(\theta_0)]),\nonumber
\end{align}
where $(\alpha,\beta)\in[0,1]^2$ are respectively related to the various possible values of $R(\theta_1)$ and $R(\theta_2)$ in (\ref{eq_C}). 
Fig. \ref{cost_ft_valley} represents $f_{\alpha,\beta}(\theta_0)$ taking the sigmoid (\ref{eqn_grad3}) for $R$, for various values of $\lambda$ and $\theta_0\in[-0.6\pi,2.6\pi]$ (in coherence with the choice done in Fig. \ref{fig_R}).
Large $\lambda$ values ($\lambda =30$, top row in Fig. \ref{cost_ft_valley}) lead to a scheme similar to the original LogQ scheme (\ref{eqn_grad2}) with a vanishing gradient in most of the domain, because the constraint (\ref{con_3}) is not satisfied which is visible in Fig.~\ref{fig_R} (blue line).
Small $\lambda$ values ($\lambda=0.5$, middle row in Fig. \ref{cost_ft_valley}) lead to a $f_{\alpha,\beta}(\theta_0)$ whose gradient does not vanish in many configurations within the considered interval, 
which means the cost function is not ensured to reach the MaxCut optimum. Indeed, the black line in Fig. \ref{fig_R} highlights that the constraint (\ref{con_1}) is not satisfied.

Intermediate $\lambda$ values ($\lambda =5$, bottom row in Fig. \ref{cost_ft_valley}) are better behaved.
The red ($\alpha=0$) and green ($\alpha=1$) curves can be considered local-minima-free.
The blue ($\alpha=0.25$) and yellow ($\alpha=0.75$) curves have a local minimum with a 'small barrier' above which carefully designed gradient (or local optimization) 'perturbations' should be able to 'jump' to reach the global minimum. 
However, the black dotted ($\alpha=0.5$) curve has a local minimum with a much higher 'barrier' that would trap gradient-inspired (or local) optimizations especially when $\beta=0.5$.
This seems problematic but in fact it is not. 
Indeed, the case $\alpha\approx\beta\approx 0.5$ 
correspond to the situation where $R(\theta_1)\approx R(\theta_2)\approx0.5$.
From Fig. \ref{fig_R}, we see that this situation is related to the largest $R'(\theta_1)$ and $R'(\theta_2)$ values 
and thus to larger gradients of $C(\theta)$. Therefore, if $\theta_1$ and $\theta_2$ reach values such that $R(\theta_1)\approx R(\theta_2)\approx0.5$ during the optimization, they will largely move away from these values in the next iterations, which will naturally lead the cost function valley to becomes more gradient (or local) optimization friendly in the $\theta_0$-direction. 

Despite these nice features, taking the sigmoid (\ref{eqn_grad3}) for $R$ is not fully compliant with the constraint (\ref{con_3})
as visible in Fig. \ref{fig_R} (green line). So, vanishing gradients can trap $\theta_0$ on the plateau related to the cost function maximum.
We tested that reducing $\lambda$ to try to mitigate this pathology is sub-optimum, rapidly leading to the other pathology discussed for the top row of Fig. \ref{cost_ft_valley}. 

\begin{figure}[h] 
    \centering
    \includegraphics[scale=0.18]{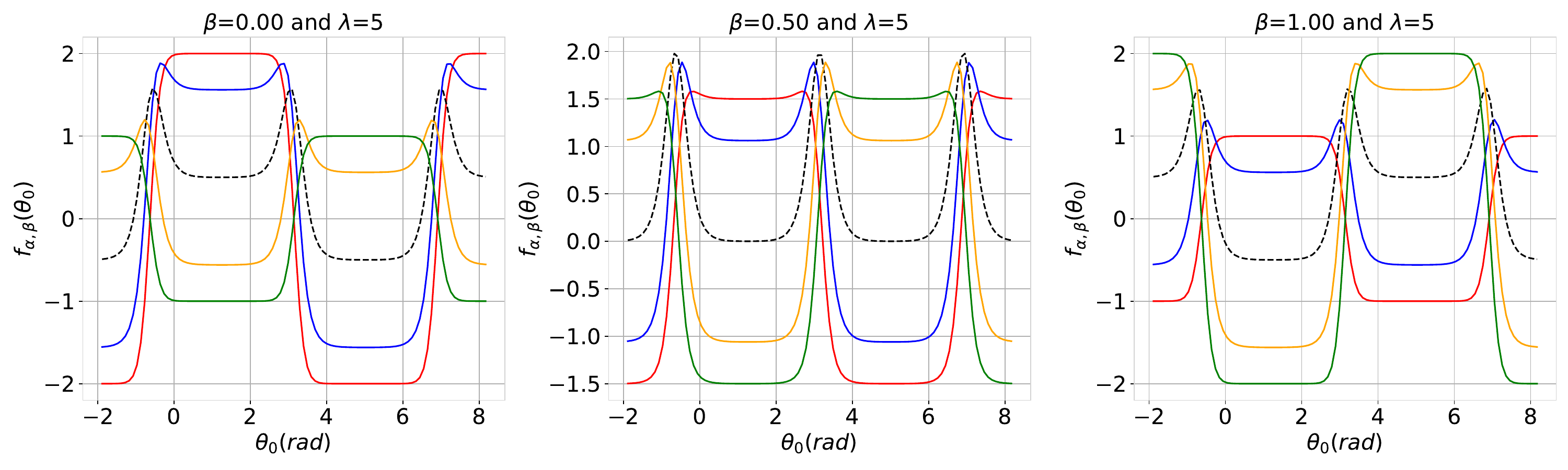}
    \caption{Representation of $f_{\alpha,\beta}(\theta_0)$ using $R$ given by (\ref{eqn_grad4}) for $\lambda=5$, $\kappa=0.2$, and $\theta_0\in[-0.6\pi,2.6\pi]$. Colors: different $\alpha$ values (similar than in Fig. \ref{cost_ft_valley}). Columns: different $\beta$ values.}
    \label{cost_ft_valley2}
\end{figure}

Fig. \ref{cost_ft_valley2} represents $f_{\alpha,\beta}(\theta_0)$ taking the 'distorted' sigmoid (\ref{eqn_grad4}) for $R$ with $\kappa=0.2$ and $\lambda=5$.
We observe that the behavior is reasonable regarding local minima (like for the bottom row of Fig. \ref{cost_ft_valley})
as well as vanishing gradient (the later are less 'spread' so that $\theta_0$ has less chance to get trapped in a maximum configuration).
This choice thus reasonably satisfies the various constraints on $R$ that we have developed
and should lead to a better performance.
Note that the extension to the $[-0.6\pi,2.6\pi]$ interval helps gradient-inspired schemes to escape more easily from local minima (reducing the 'path-distance' to escape a local minimum valley).

\section{Application to larger problems}

We use (\ref{eqn_grad4}) for $R$ with $\kappa=0.2$ and update the parameters in $[-0.6\pi,2.6\pi]$ only. 
We use Cobyla for the optimization, start the iterations with $\lambda=5-6$ and a large 'rhobeg' parameter ($\approx 3$ to help to escape small 'barriers' local minima and small plateaus), and decrease rhobeg as the iterations progress. Once converged, we do few final iterations with $\lambda=30$ as a 'post-processing'. 
We implement a 'multistart' approach which explores a few random initializations of $\theta$ in $[0,2\pi]^n$ and keep the one with the smallest $C(\theta)$ to initialize the gradient-inspired (or local optimization) iterations, which helps efficiency.
The scheme has proven to be robust in many configurations, and we call it LogQ-grad in the following (to differentiate it from the original LogQ scheme that uses GA).

We present results obtained for the MaxCut problem.
Graphs were generated using Python's $fast\_gnp\_random\_graph$ function with a density of $0.3$ and $seed=0$ (we verified our conclusions are robust for different seeds).
Our tests were done using the quantum simulator \textit{statevector\_simulator}, from Eviden as well as IBM.

Previous work already discussed the advantage of original LogQ over QAOA on the MaxCut problem \cite{chatterjee2023solving}.
We did various tests which confirmed these advantage. For instance, for $n=16$ we obtained equivalent results between QAOA (with one layer, $p=1$) and original LogQ, but QAOA's runtime was an order of magnitude longer than LogQ's runtime.
The difference in time increases for larger instances.
QAOA tends to become very costly to be simulated when $n > 30$.

\begin{figure}[ht]
    \centering    
        \includegraphics[scale=0.55]{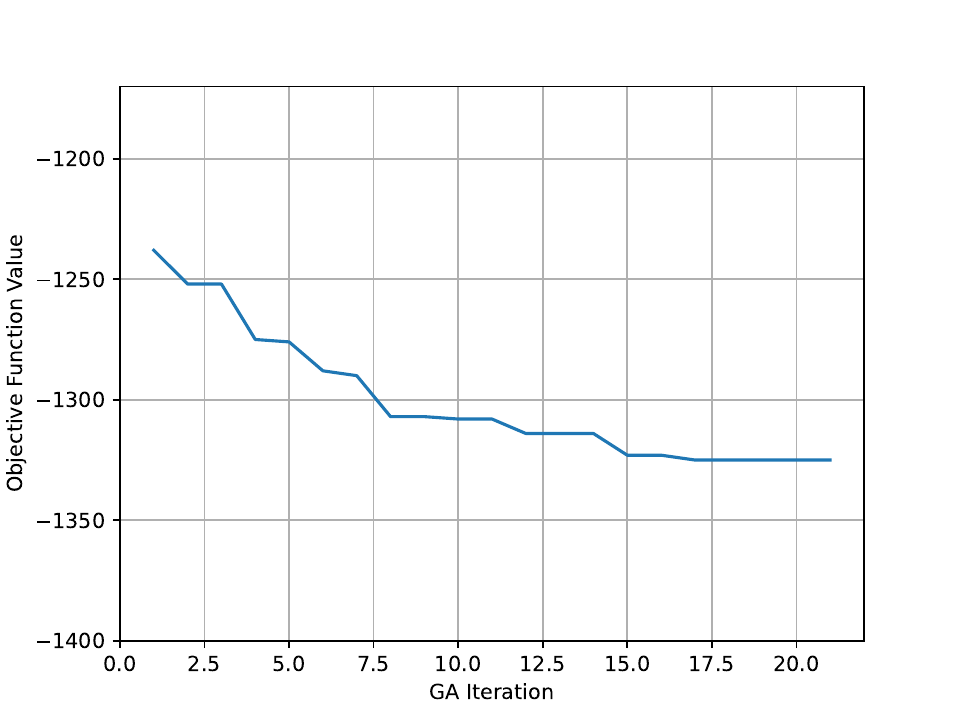}
        \includegraphics[scale=0.55]{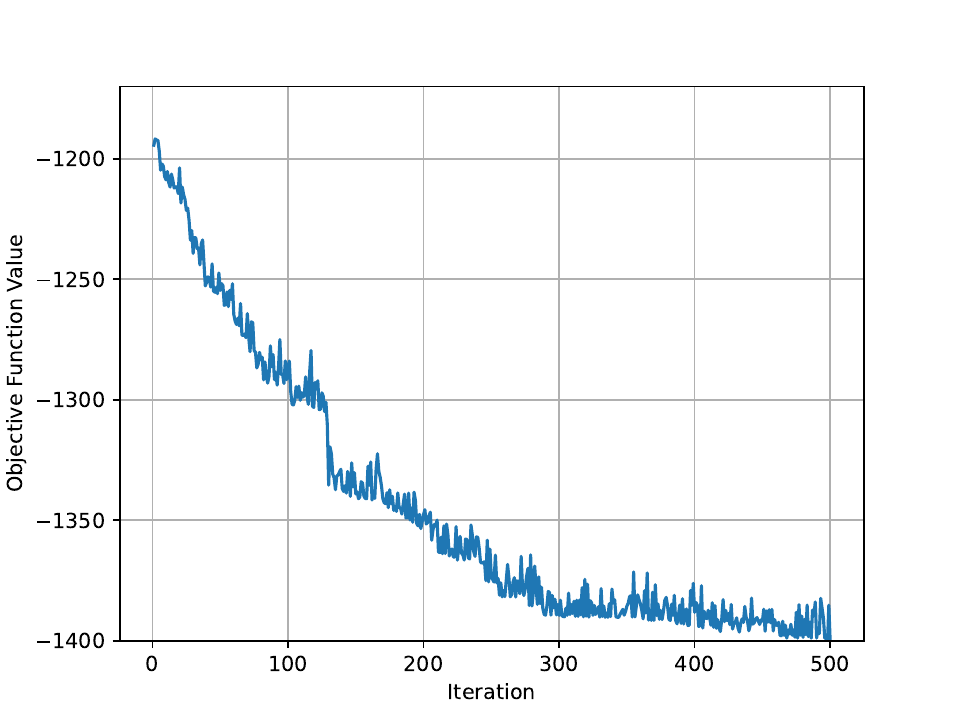}
    \caption{MaxCut results for $n=128$ vertices ($N=7$ LogQ qubits): Cost function value evolution during parameters optimization for original LogQ (using a GA, top) and LogQ-grad (using Cobyla, bottom).}
    \label{fig:results}
\end{figure}

In the following, we focus on comparing original LogQ (using GA) and LogQ-grad on instances that are too big to be simulated for QAOA ($n=50, 128, 256$).
We tuned the GA optimization (population and iterations) so that the runtime of original LogQ is comparable to the one of LogQ-grad and kept the best GA results over many runs (as GA is non-reproducible). This allowed us to focus only on accuracy improvement when using LogQ-grad.
%
Table \ref{LogQ-grad} and Figure  \ref{fig:results} present obtained results.
We observe LogQ-grad brings a systematic accuracy improvement  (lower cost function value) in addition to be reproducible, which is satisfying.
Importantly, in all cases LogQ-grad leads after convergence to $R(\theta^*)=0$ or $1$ (only), in agreement with the theory we developed. 

Note that the number of GA iterations is not directly comparable to the number of LogQ-grad iterations. This is because for every GA iteration (also called GA 'generation'), the objective function is calculated up to 'population size' times. The true number of objective function calls for GA is therefore \textit{population\_size} * \textit{GA\_iterations}.
Also, we remind these experiments were performed on quantum simulators which is sufficient to conclude on the interest of the LogQ-grad scheme. An evaluation of the original LogQ algorithm on quantum computers can be found in \cite{chatterjee2023solving}.
%

\begin{table}[htbp]
\caption{Comparison between LogQ-grad and original LogQ cost function values for MaxCut ($n=$ number of graph vertices, $N=$ number of LogQ qubits). The runtimes between the two methods are comparable for the same instance.
'it.' and 'pop.' represent GA iterations and population size respectively. }
\centering
    \begin{tabular}{|c|c|c|}
        \hline
         & LogQ-grad & Orig. LogQ (GA)  \\
        \hline
        MaxCut $n=50$, $N=5$ & -238 (it. 250) & -219 (it. 20, pop. 20) \\
        \hline
        MaxCut $n=128$, $N=7$ & -1410 (it. 500) & -1325 (it. 20, pop. 25)  \\
        \hline
        MaxCut $n=256$, $N=8$ & -5383 (it. 750) & -5149 (it. 30, pop. 25)  \\
        \hline
    \end{tabular}
    \label{LogQ-grad}
\end{table}

%

\section{Conclusion}

While LogQ gives a way to encode QUBO problems with exponentially fewer qubits than QAOA, 
its optimization remained a challenge and hampered the efficiency of the scheme.
We justified a new LogQ parameterization that can be optimized with an efficient gradient-inspired method.
We illustrated the scheme on an analytical model and presented larger scale  MaxCut results.
The scheme brings a systematic improvement over original LogQ and thus represents a step towards solving large QUBO problems using quantum algorithms.
Our research continues, exploring other LogQ parameterizations than (\ref{eqn_grad4}) that meet the conditions we developed for $R$,
and better quantifying the '$large$', '$small$' and '$close$' in (\ref{con_3})-(\ref{con_4}).

The main remaining challenge is to reduce the complexity of the decomposition (\ref{L_decompo}) of the operator $\hat{L}$ and thus of the measurements, which  scales maximally as $(n^2+n)/2$. 
This does not hamper the efficiency of LogQ in situations where the number of QUBO parameters $n$ is $\le 10^4$, but things become more challenging when $n > 10^4$. 
For instance, in the $n=10^5$ binary variables case, LogQ requires less than $N=20$ qubits and a quantum circuit of $n=10^5$ CNOTs, 
which would be very efficient if we could find a way not to compute all of the $(n^2+n)/2\approx 5.10^9$ tensor products $J_k$ in (\ref{L_decompo}).
We are working on a method to \textit{a priori} define the terms for which the $Tr(J_k L)$ take negligible values, from symmetries and properties of the considered application. Another direction is to work on other decompositions than (\ref{L_decompo}) that would be sparse by nature. 

These research will allow us to scale applications of LogQ to industrially relevant QUBO problems, portfolio optimization being one of our primary target.

 \section*{Acknowledgment}

We are grateful to Henri Calandra for insightful discussions and TotalEnergies for the permission to publish this work.

\bibliography{conference_101719}

\providecommand{\noopsort}[1]{}\providecommand{\singleletter}[1]{#1}
\begin{thebibliography}{10}

\bibitem{qubotutorial}
F.~Glover, G.~Kochenberger, and Y.~Du, ``Quantum bridge analytics i: a tutorial on formulating and using qubo models,'' {\em 4OR, Springer}, vol.~17(4), pp.~335--371, 2019.

\bibitem{date2021qubo}
P.~Date, D.~Arthur, and L.~Pusey-Nazzaro, ``Qubo formulations for training machine learning models,'' {\em Scientific reports}, vol.~11, no.~1, p.~10029, 2021.

\bibitem{calude2017qubo}
C.~S. Calude, M.~J. Dinneen, and R.~Hua, ``Qubo formulations for the graph isomorphism problem and related problems,'' {\em Theoretical Computer Science}, vol.~701, pp.~54--69, 2017.

\bibitem{papalitsas2019qubo}
C.~Papalitsas, T.~Andronikos, K.~Giannakis, G.~Theocharopoulou, and S.~Fanarioti, ``A qubo model for the traveling salesman problem with time windows,'' {\em Algorithms}, vol.~12, no.~11, p.~224, 2019.

\bibitem{karp}
R.~M. Karp, ``Reducibility among combinatorial problems,'' {\em In: Miller, R.E., Thatcher, J.W., Bohlinger, J.D. (eds) Complexity of Computer Computations. The IBM Research Symposia Series. Springer, Boston, MA.}, 1972.

\bibitem{surveypoly}
J.~A. \relax Ruiz-Vanoye~et al., ``Survey of polynomial transformations between np-complete problems,'' {\em Journal of Computational and Applied Mathematics}, vol.~235, pp.~4851--4865, 2011.

\bibitem{cerezo2021variational}
M.~Cerezo, A.~Arrasmith, R.~Babbush, S.~C. Benjamin, S.~Endo, K.~Fujii, J.~R. McClean, K.~Mitarai, X.~Yuan, L.~Cincio, {\em et~al.}, ``Variational quantum algorithms,'' {\em Nature Reviews Physics}, vol.~3, no.~9, pp.~625--644, 2021.

\bibitem{peruzzo2014variational}
A.~Peruzzo, J.~McClean, P.~Shadbolt, M.-H. Yung, X.-Q. Zhou, P.~J. Love, A.~Aspuru-Guzik, and J.~L. O’brien, ``A variational eigenvalue solver on a photonic quantum processor,'' {\em Nature communications}, vol.~5, no.~1, pp.~1--7, 2014.

\bibitem{moll2018quantum}
N.~Moll, P.~Barkoutsos, L.~S. Bishop, J.~M. Chow, A.~Cross, D.~J. Egger, S.~Filipp, A.~Fuhrer, J.~M. Gambetta, M.~Ganzhorn, {\em et~al.}, ``Quantum optimization using variational algorithms on near-term quantum devices,'' {\em Quantum Science and Technology}, vol.~3, no.~3, p.~030503, 2018.

\bibitem{qaoa}
E.~Farhi, J.~Goldstone, and S.~Gutmann, ``A quantum approximate optimization algorithm,'' {\em arXiv preprint arXiv:1411.4028}, 2014.

\bibitem{qaoa2}
S.~Hadfield, Z.~Wang, B.~O’Gorman, E.~G. Rieffel, D.~Venturelli, and R.~Biswas, ``From the quantum approximate optimization algorithm to a quantum alternating operator ansatz,'' {\em Algorithms}, vol.~12, no.~2, 2019.

\bibitem{rancic}
M.~J. Ran{\v{c}}i{\'c}, ``Noisy intermediate-scale quantum computing algorithm for solving an n-vertex maxcut problem with log (n) qubits,'' {\em Physical Review Research}, vol.~5, no.~1, p.~L012021, 2023.

\bibitem{chatterjee2023solving}
Y.~Chatterjee, E.~Bourreau, and M.~J. Ran{\v{c}}i{\'c}, ``Solving various np-hard problems using exponentially fewer qubits on a quantum computer,'' {\em Phys. Rev. A}, vol.~109, p.~052441, 2024.

\bibitem{chatterjee2023hybrid}
Y.~Chatterjee, Z.~Allybokus, M.~J. Ran{\v{c}}i{\'c}, and E.~Bourreau, ``A hybrid quantum-assisted column generation algorithm for the fleet conversion problem,'' {\em arXiv:2309.08267}, 2023.

\bibitem{neasqc}
Y.~Chatterjee, E.~Bourreau, and H.~Calandra, ``D5.9: Benchmarking of qaoa-based algorithms for mesh segmentation, against k-means, normalized and randomized cuts and core extraction methods,'' 2024.

\bibitem{brandhofer2022benchmarking}
S.~Brandhofer, D.~Braun, V.~Dehn, G.~Hellstern, M.~H{\"u}ls, Y.~Ji, I.~Polian, A.~S. Bhatia, and T.~Wellens, ``Benchmarking the performance of portfolio optimization with qaoa,'' {\em Quantum Information Processing}, vol.~22, no.~1, p.~25, 2022.

\bibitem{fuchs2021efficient}
F.~G. Fuchs, H.~{\O}. Kolden, N.~H. Aase, and G.~Sartor, ``Efficient encoding of the weighted max k-cut on a quantum computer using qaoa,'' {\em SN Computer Science}, vol.~2, no.~2, pp.~1--14, 2021.

\bibitem{rath2023quantumdataencodingcomparative}
M.~Rath and H.~Date, ``Quantum data encoding: A comparative analysis of classical-to-quantum mapping techniques and their impact on machine learning accuracy,'' 2024.

\bibitem{yang2018efficient}
X.~Yang, J.~Liu, Y.~Yang, Q.~Qing, and G.~Wen, ``An efficient topology description function method based on modified sigmoid function,'' {\em Mathematical Problems in Engineering}, vol.~2018, no.~1, p.~3653817, 2018.

\end{thebibliography}

\end{document}